\documentclass[a4paper,11pt]{article}

\usepackage{latexsym}
\usepackage{amsfonts}
\usepackage{color}
\usepackage{bbm}
\usepackage{amsmath}
\usepackage{amssymb}
\usepackage{tabularx}
\usepackage{chicagoz}
\usepackage{theorem}
\usepackage{pstricks,pst-plot}
\usepackage{graphicx}
\usepackage{epsfig}
\usepackage{lscape}
\usepackage{rotating}
\usepackage{float}
\usepackage{cite}
\usepackage{caption}
\usepackage{tabulary}
\usepackage{hyperref}

\author{Eduardo Ramos-P\'erez$^{(1)}$,\\ Pablo J. Alonso-Gonz\'alez$^{(2)}$, Jos\'e Javier N\'u\~nez-Vel\'azquez$^{(2)}$ }
\title{Stochastic reserving with a stacked model based on a hybridized Artificial Neural Network}

\title{Stochastic reserving with a stacked model based on a hybridized Artificial Neural Network}

\date{\footnotesize\textit{(1) Ph D Student (Economics and Management Program). Universidad de Alcal\'a.\\
	(2) Economics Department. Universidad de Alcal\'a.}
\thanks{\hskip -1.8em
   Authors' address: (1)\&(2) :Economics Department, Universidad de Alcal\'a, Plaza de la Victoria 2, 28802 Alcal\'a de Henares, Spain.
   E--mails: P.J. Alonso-Gonz\'alez, \texttt{pablo.alonsog@uah.es}, J.J. N\'u\~nez, \texttt{josej.nunez@uah.es}, E. Ramos,  \texttt{ramos.perez.e@gmail.com}}
\thanks{\hskip -1.8em Corresponding author: P.\, Alonso;\,\, Date: \today. This manuscript version is made available under the CC-BY-NC-ND 4.0 license http://creativecommons.org/licenses/by-nc-nd/4.0/}
}

\begin{document}
\maketitle
\begin{abstract}
\noindent
Currently, legal requirements demand that insurance companies increase their emphasis on monitoring the risks linked to the underwriting and asset management activities. Regarding underwriting risks, the main uncertainties that insurers must manage are related to the premium sufficiency to cover future claims and the adequacy of the current reserves to pay outstanding claims. Both risks are calibrated using stochastic models due to their nature. This paper introduces a reserving model based on a set of machine learning techniques such as Gradient Boosting, Random Forest and Artificial Neural Networks. These algorithms and other widely used reserving models are stacked to predict the shape of the runoff. To compute the deviation around a former prediction, a log-normal approach is combined with the suggested model. The empirical results demonstrate that the proposed methodology can be used to improve the performance of the traditional reserving techniques based on Bayesian statistics and a Chain Ladder, leading to a more accurate assessment of the reserving risk.
\end{abstract}
{\small
}

\section{Introduction}
\label{introd}
As with any other company, the survival of an insurance firm depends on its ability to obtain a sustainable profit over the years. These entities have to offer their services at an adequate and competitive premium, while the ultimate cost of the claims is subject to uncertainty. Thus, reserving models were developed in order to estimate and monitor the expected ultimate cost of outstanding claims. Although life insurance contracts manifest uncertainty about the claims cost, reserving takes a special relevance in general insurance as that uncertainty tends to be higher, at least in the short term.\\

Methods of estimating the level of reserves in non-life insurance have evolved from classical and deterministic methods toward others that take into account the loss reserve uncertainty. The aim of the first type is to estimate the expected level of reserves by taking the historical information into consideration. Chain Ladder is the most frequently used method of this family. When historical data are not stable enough to use the Chain Ladder technique, the \citeN{BF_1972} model tends to be the preferred option to obtain an adequate estimate of the expected ultimate cost.\\

The increasing interest of investors in the risk profile of financial institutions since the Financial Crisis of 2007-2008 and the implementation of the Solvency II Directive in the European market have fostered the use of stochastic reserving models. As in the case of deterministic approaches, stochastic models based on the Chain Ladder technique are the most commonly used. One of the main techniques within this family is the Overdispersed Poisson (ODP) model developed by \citeN{Renshaw_1998} and its bootstrap implementation suggested by \citeN{england_verrall_1999} and \citeN{england_2002} which assumes that incremental claims follow an ODP distribution where the variance is proportional to the mean. \\

In this model, incremental claims must be positive, but this limitation can be overcome by using the quasi-likelihood approach introduced by \citeN{mccullagh_nelder_1989}. In cases where the ODP assumption does not properly fit the data, \citeN{Kremer_1982}, \citeN{Mack_1991} and \citeN{Verrall_2000} developed other models assuming log-normal, gamma and negative binomial distributions respectively. In contrast to the methods within this family, \cite{Mack_1993} developed a free-distribution model by focusing and limiting the claims reserve distribution analysis to the first two moments.\\ 

Thus, the bootstrap implementation of Mack's model allows the analyst to obtain a reserve distribution without the necessity of defining a theoretical distribution for the cumulative or incremental claim cost. If the bootstrapping procedure is to be avoided, \citeN{england_verrall_2006} introduced a stochastic Bayesian implementation of the ODP, Negative Binomial and this last free-distribution model. This approach was recently expanded by \citeN{Meyers_2015}, who developed some Bayesian Markov Chain Monte-Carlo (MCMC) models (Levelled Chain-Ladder, Correlated Chain-Ladder, Levelled Incremental Trend, Correlated Incremental Trend and Changing Settlement Rate) for incurred and paid data. Their aim is to improve the performance of ODP and Mack models by using different approaches such as recognizing the correlation between accident years, including a skewed distribution to model negative incremental payments, introducing a trend over the development years and allowing changes in the claim settlement rate.\\

Another set of models is focused on using several triangles simultaneously in order to take into consideration different characteristics of incurred and paid data. The main models within this family are the Munich Chain Ladder (MCL) method and Double Chain Ladder (DCL) model developed by \citeN{QM_2004} and \shortciteN{MNV_2012}, respectively. By modifying this last method, \shortciteN{MEV_2018} addressed the problem of calculating general insurance reserves when the portfolio is covered by an excess-of-loss reinsurance. In addition to MCL and DCL, \citeN{MW_2010} introduced a Bayesian implementation of the paid-incurred chain (PIC) reserving method (\shortciteNP{PCV_2008}) based on using both incurred and paid data. \shortciteN{HMW_2012} and \citeN{HW_2013} also investigated and developed models related to the PIC method, while \citeN{Hal_2009} and \citeN{Ven_2008} introduced regression approaches based on using both data sources. \shortciteN{PAD_2014}, \citeN{AP_2014}, and \shortciteN{MNV_2013} also proposed models by taking into consideration different data sources to estimate the expected ultimate claim cost.\\

In addition to the different approaches exposed above, it is possible to find models where the information is not organized in an aggregated way, as in the classical triangles, but rather in individual claims data (see \shortciteNP{TMS_2008}, \shortciteNP{JMS_2011}, \shortciteNP{PAD_2013}, \citeNP{AP_2014}, \shortciteNP{MNV_2015}, \citeNP{CP_2016}, or \citeNP{W_2018}).\\

Thanks to the increase in computational power, machine learning techniques have turned into an adequate tool for reserving purposes. Artificial Neural Networks (\citeNP{GW_2018} and \citeNP{W_2018}), regression trees (\citeNP{W2_2018}), Recurrent Neural Networks (\citeNP{Kuo_2018}) or tree-based algorithms (\shortciteNP{LMT_2019}) have been used to predict claim reserves. \shortciteN{GRW_2018} embedded the ODP model into a neural network framework, and \citeN{BR_2019} introduced a nonparametric reserving model based on extremely randomized trees (\shortciteNP{GEW_2006}) and individual claims data. In addition to the aforementioned algorithms, other machine learning techniques were used by \shortciteN{MNV2_2013} for reserving purposes, and a support vector machine was applied to classify risks prior to the reserve calculation (\shortciteNP{DTM_2011}).\\

The research carried out in this paper develops a nonparametric reserving model based on the stacking algorithm methodology. The proposed architecture consists of two different levels. Random Forest (RF) (\citeNP{Breiman_2001}), Gradient Boosting (GB) with regression trees (\shortciteNP{Friedman_2000}), Artificial Neural Network (ANN) (\citeNP{Mcculloch_1943}), Changing Settlement Rate (CSR) reserving model and the Chain Ladder assumptions are incorporated within the first level, while an ANN is included in the second level of the stacked model (Stacked-ANN) architecture in order to generate the final predictions. Therefore, the aim of this hybrid model is to improve the performance of the individual components by creating an architecture that can to learn from the different algorithms and the reserving models included within the first level. \\

Although the overall methodology is based on that proposed by \shortciteN{RAN_2019} for stock volatility forecasting purposes, the model architecture proposed in this study is different. In this research, machine learning algorithms and reserving models are present in the first level, while in the architecture developed by \shortciteN{RAN_2019}, only machine learning algorithms were included. Therefore, the most popular models for forecasting volatility such as GARCH or EGARCH were not integrated within the model architecture, while in this case, Chain Ladder and CSR are incorporated. It is also worth mentioning that in contrast to the hybrid model proposed for forecasting volatility purposes, in this research, the second level only receives information already processed by the models within the first level. In addition to the main differences explained above, it should be pointed out that the stacking algorithm methodology has not appeared previously in the actuarial literature related to the valuation of loss reserves. Apart from that, a log-normal approach is combined with the suggested reserving model based on machine learning in order to compute the reserve variability.\\

As all the different algorithms and reserving models of the first level are incorporated in the ANN of the second level, some of the most important research studies carried out in the context of selecting the optimal ANN architecture will be discussed. There is a significant amount of literature supporting the use of ANNs with just one hidden layer because under mild assumptions on the activation functions, the universal approximation theorem states that a feedforward ANN with a single hidden layer and a finite number of neurons can approximate any continuous function on compact subsets of the Euclidean space. \\

Based on regularization techniques and using just one hidden layer network, \citeN{PG_1990} developed a theoretical framework to approximate nonlinear mappings named regularization networks. These authors demonstrated that their architecture can approximate any continuous function on a compact domain if the number of units is high enough. \citeN{Cyb_1989} and \shortciteN{HSW_1989} also proved that one hidden layer networks with sigmoidal activation functions can approximate continuous functions on any compact Euclidean space. It was also shown that, under certain conditions, an arbitrarily small error between a single hidden layer ANN and any other continuous function can be obtained by increasing the number of neurons (\citeNP{Bar_1994}, \citeNP{FUN_1989} and \citeNP{Hor_1993}). \citeN{Nak_2011} showed that the range of effective learning rates is wider in the case of ANN with one hidden layer than in architectures with multiple hidden layers.\\

On the other hand, \citeN{Hor_1991} and \shortciteN{LLP_1993} demonstrated that ANNs have the potential of being universal approximators not only due to the choice of a specific activation function but also because of the possibility of using several hidden layers. Limitations of the approximation capabilities of one hidden layer networks were demonstrated by \shortciteN{CLM_1994} and \shortciteN{CLM_1996}. In recent years, multi-hidden layer architectures have improved the state of the art in machine learning. \\

For example, in the context of natural language processing, the models and architectures created by \shortciteN{BERT_2018} (BERT), \shortciteN{GPT3_2020} (GPT3) and \shortciteN{VSP_2017} (Transformer) overcome the performance of other less complex models. In addition, it is worth mentioning that agents trained with multi-hidden layer ANNs have been able to overcome the human performance in specific tasks such as playing chess (\shortciteNP{SHS_2017}) or `go' (\shortciteNP{SHA_2016}). With respect to the optimal number of neurons, \citeN{C_2007} analysed this issue in the context of solving the dynamic network loading problem, while \citeN{SD_2013} proposed a list of principles to select this number.\\

Results from recently published papers in the actuarial field support the idea of applying ANNs with multiple hidden layers. Indeed, \citeN{RW_2018} and \shortciteN{NLM_2019} applied this structure to model human mortality, while \shortciteN{CFM_2018} used it for estimating the economic capital of insurance companies under the Solvency II framework. Thus, the ANNs included within the architecture of the Stacked-ANN model have several hidden layers.\\

The rest of the paper proceeds as follows: Section \ref{Benchmark} presents the set of models used for comparison purposes. Additionally, the error and risk measures taken to validate the stochastic reserves, payments and ultimate losses are discussed. In Section \ref{stack}, the theoretical background and architecture of the reserving model based on stacking algorithms (Stacked-ANN) are explained. Details about the log-normal approach proposed for obtaining a stochastic distribution are also given in this section. The empirical results, error and risk measures of the different reserving models are shown in Section \ref{resul}. Finally, Section \ref{conc} presents the main conclusions derived from the results and comparisons presented in Section \ref{resul}.

\section{Benchmark models and validation}
\label{Benchmark}
As previously stated, this section explains the benchmark models and the different measures used to assess their performance. Thus, the first paragraphs are dedicated to ODP, Mack's model, CSR and a nonparametric approach based on ANNs, while the end of this section presents the indicators used to compare and validate the reserve distribution functions estimated by the benchmark models with those simulated by the model presented in Section \ref{stack}.\\

The first benchmark model is ODP (\citeNP{Renshaw_1998} and \citeNP{england_verrall_1999}). Denoting the origin year as \(i\) and the development year as \(j\), this reserving model based on the Chain Ladder technique assumes that incremental payments, \(C_{ij}\), follow an overdispersed Poisson distribution with a variance proportional to the mean:
\begin{align}
&E[C_{ij}]=\mu_{ij}     &Var[C_{ij}]= \phi\mu_{ij}
\end{align}
where \(\phi\) is the parameter that determines the level of overdispersion. Even though this model assumes \(C_{ij}\) to be a positive integer, the quasi-likelihood (\citeNP{mccullagh_nelder_1989}) approach allows fits the model to non-integer data, which can be either positive or negative. The bootstrapping procedure used in this study to compute a reserve distribution function with the ODP model was introduced by \citeN{england_verrall_1999} and \citeN{england_2002}.\\

\citeN{Mack_1993} model, which is also based on the Chain Ladder technique, is the second benchmark. The main characteristic of this reserving model is the lack of assumptions about the underlying distribution of the payments. This is achieved by using only the first two moments:
\begin{align}
&E[D_{ij}]=\lambda_{j} D_{i,j-1}     &Var[D_{ij}]= \sigma^2_j D_{i,j-1}
\end{align}
where \(\lambda_{j}\) and \(\sigma^2_j\) refer to the parameters to be estimated, and \(D_{ij}\) is the cumulative payment. As with the ODP model, a bootstrapping procedure is used to calculate the reserve distribution function with Mack's model.\\

The third benchmark model is CSR, a Bayesian approach introduced by \citeN{Meyers_2015}. The default calibration and prior distributions suggested by this author will be used in this study:
\begin{itemize}
\item \(\alpha_i \sim N(\ln{P_i}+logelr,\sqrt{10})\), where \(logelr \sim U(-1,0.5)\) and \(P_i\) are the premiums by accident year.
\item \(\beta_j \sim U(-5,5)\) for \(j=1,...,J-1\). In the last development year, \(\beta_J=0\).
\item \(\mu_{i,j}=\alpha_i+\beta_j(1-\gamma)^{i-1}\), where \(\gamma \sim N(0,0.025)\).
\item Each \(\sigma_j=\sum_{i=j}^{J}{a_i}\), where \(a_i \sim U(0,1)\).
\end{itemize}
Taking into consideration the aforementioned distributions and parameters, the cumulative payments simulated by the CSR model follow a log-normal distribution, \(D_{i,j} \sim LN(\mu_{i,j},\sigma_j)\), subject to the constraint \(\sigma_1 > \sigma_2 > ... > \sigma_J\).\\

To analyse the improvement in the performance due to the stacking procedure that is presented in Section \ref{stack}, the last benchmark model to be introduced is an individual ANN. The inputs and characteristics (hidden layers, activation functions, etc.) of this algorithm will be the same as those used for the ANN included within the first level of the Stacked-ANN. Additionally, the log-normal procedure to obtain the reserve variability is the same as that for the Stacked-ANN model. To avoid repeating content, refer to Section \ref{stack} for further details about the characteristics of the ANN used as a benchmark.\\

Once the four benchmark models are explained, the different measures selected to compare the performance of the Stacked-ANN with the aforementioned reserving models are presented. Insurance regulations such as the Solvency II Directive and Swiss Solvency Test ask the general insurance companies to evaluate their expected reserves and potential deviations from these central scenarios. Thus, the error of the estimated reserves will be computed in order to compare the performance of the different models. As several triangles with different levels of payments are used during this study, the measure for evaluating the reserves is
\begin{align}
\%RMSE(R^{t})=\frac{\sqrt{\sum^{K}_{k=1}{(\hat{R}^{t}_{k,\mu}-R^{t}_k)^2}/K}}{\sum^{K}_{k=1}{R^{t}_k}}*100=\frac{RMSE(R^{t})}{\sum^{K}_{k=1}{R^{t}_k}}*100
\end{align}
where \(K\) is the total number of triangles, \(t\) is the calendar year when the reserves are evaluated, \(\hat{R}^{t}_{k,\mu}\) is the reserve predicted by the reserving model using the triangle \(k\) and \(R^{t}_k\) the reserves that were actually observed for that triangle. As it can be derived from the former expression, the aim of this error measure is the evaluation of the weight of the root mean squared error over the total reserves. To understand the model's performance, this error measure will also be calculated for the next year's payments (\(\%RMSE(P^{t+1})\)) and the ultimate loss cost (\(\%RMSE(U^{t})\)).\\

In addition to the aforementioned error measures, the reserving risk (\(RR\)) per unit of reserve derived from the use of the different stochastic reserving models will be analysed. As previously stated, the models are going to be fitted to several triangles, so the average of the former ratio is taken as a risk measure:
\begin{align}
Ratio(RR^{t}_{1-\alpha})=\frac{\sum^{K}_{k=1}{(\hat{R}^{t}_{k,1-\alpha}-\hat{R}^{t}_{k,\mu})/\hat{R}^{t}_{k,\mu}}}{K}=\frac{\sum^{K}_{k=1}{RR^{t}_{1-\alpha}/\hat{R}^{t}_{k,\mu}}}{K}
\end{align}
where \(\hat{R}^{t}_{k,\mu}\) is the mean and \(\hat{R}^{t}_{k,1-\alpha}\) is the percentile \(1-\alpha\) of the estimated reserve distribution function of the company \(k\). A deeper evaluation of the variation estimated by the different stochastic models is carried out by calculating the standard deviation per unit of reserve:
\begin{align}
Ratio(\sigma)=\frac{\sum^{K}_{k=1}{\sigma(\hat{R}^{t}_k)/\hat{R}^{t}_{k,\mu}}}{K}
\end{align}
Finally, in order to check the adequacy of the reserving risk calculated for the different companies, the \citeN{Kupiec_1995} test is applied in order to verify if the number of excesses is aligned with the selected confidence level. The empirical results of the test and measures are collected in Section \ref{resul}.

\section{Stochastic reserving model based on the stacking algorithm approach}
\label{stack}
\begin{figure}[!htb]
\begin{center}
\caption{Stacked-ANN model structure}
\includegraphics[width=0.99\textwidth]{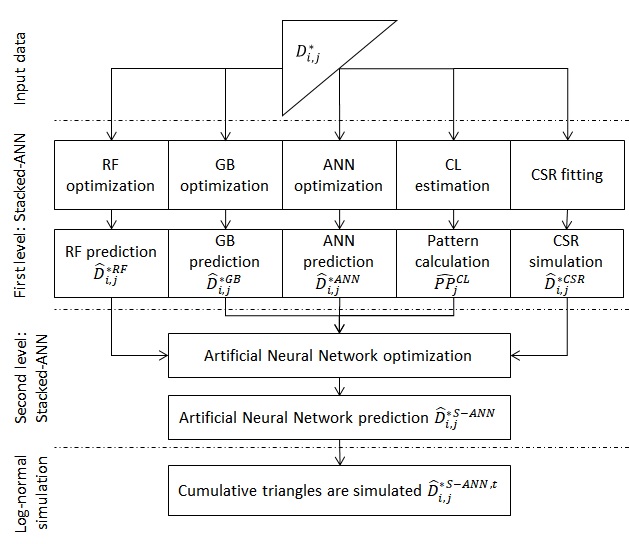}
\label{fig:Fig 1}
\end{center}
\end{figure}

This section is divided into several subsections in order to sequentially explain the proposed reserving model. In addition, Figure \ref{fig:Fig 1} presents the model architecture in order to support the explanation.

\subsection{Model inputs}
\label{stack.data}
Before estimating the different reserving models within the first level of the Stacked-ANN model, the database used, as well as the response and explicative variables for fitting the algorithms within this level, need to be defined.\\

The lower and upper triangles needed to fit and validate the models are obtained from Schedule P of the NAIC Annual Statement. This database (available on the \href{https://www.casact.org/research/index.cfm?fa=loss_reserves_data}{CAS website}) was collected from property and casualty insurers that underwrite business in the US, and it contains both paid and incurred losses (net of reinsurance) of the accident years from 1988 to 1997. Ten development years are available for every accident year. In addition to loss data, gross and net premiums by accident year are also reported in the database.\\

In this paper, the different reserving models will be fitted to 200 loss triangles from NAIC Schedule P, 50 from each of the following lines of business: Commercial Auto (CA), Private Passenger Auto Liability (PA), Workers' Compensation (WC) and Other Liability (OL). As pointed out by \citeN{Meyers_2015}, selecting triangles from insurers who made significant changes in business operations is one of the main mistakes that could be made with NAIC Schedule P data. The coefficient of variation of the net premiums and the net/gross premium ratio should be appropriate indicators of changes in business operations, so this author selected insurers that minimize the aforementioned metrics. The triangles selected by \citeN{Meyers_2015} are used in this research in order to avoid the former issue and ensure comparability with other studies.\\

With regard to the explanatory variables, as with other nonparametric reserving models based on Generalized Additive Models (\citeNP{hastie_1986} and \citeNP{england_verrall_2002}) or RNN (\citeNP{Kuo_2018}), accident \(i\) and development \(j\) years were selected to be the inputs of the first-level algorithms. Both were initialized as one and then scaled to range \([0,1]\) (hereinafter \(AY^*_i\) and \(DY^*_j\)) in order to facilitate the fitting of the algorithms (\shortciteNP{Hastie_2009}).\\

The response variable of these algorithms is the scaled cumulative payments \(D^*_{ij}\). Depending on the data availability and the characteristics of the portfolio to be modelled, different exposure measures can be selected to scale \(D_{ij}\). In this paper, net premiums \(P_{i}\) play the role of exposure measure, as this is the most relevant option between the variables available in the database.\\

\begin{figure}[!htb]
\begin{center}
\caption{Train and test sets}
\includegraphics[width=0.99\textwidth]{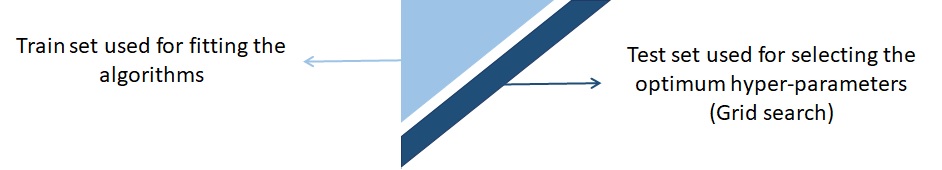}
\label{fig:Fig 3}
\end{center}
\end{figure}

Loss triangles are a representation of payments over time by accident or underwriting year. Thus, the training and optimization of the deep learning algorithms within the Stacked-ANN model architecture need to take into consideration that loss triangles are composed of temporal series. Accordingly, the last diagonal is selected as a test set because it contains the most updated information, while the rest of the triangle is used for fitting the algorithms (Figure \ref{fig:Fig 3}).\\

During the optimization process, different configurations of the algorithms are fitted with the training data. To obtain the best configuration, the test set is predicted, and the root mean squared error of every option is computed. Finally, the configuration that minimizes the former test error is selected.\\

\subsection{First level: Individual models}
\label{stack.first}
The first level of the Stacked-ANN model consists of a Chain Ladder, CSR, and three algorithms whose inputs were described in Section \ref{stack.data}. It is worth mentioning that as ODP and Mack's model are based on the Chain Ladder technique, the Stacked ANN model incorporates the core rationale behind these stochastic reserving models. The machine learning algorithms (RF, GB and ANN) fitted at this step are explained in the following paragraphs and will be optimized by applying a grid search to some hyperparameters and by measuring the test error. Additionally, at the end of this subsection, the Chain Ladder and CSR hypothesis are integrated within the Stacked-ANN model architecture.\\

The Random Forest (RF) algorithm introduced by \citeN{Breiman_2001} averages \(B\) different regression trees. In every fitted tree, the explanatory variables and data points used during the training are randomly selected. Therefore, the formal expression to predict the scaled cumulative payments is:
\begin{align}
\hat{D}^{*RF}_{ij}=\frac{\hat{D}^{RF}_{ij}}{P_{i}}=\frac{\sum^{B}_{b=1}{T_b(X)}}{B}
\end{align}
\(T_b\) represents the \textit{b-th} regression tree fitted and \(X\) the selected subset of \(AY^*_i\) and \(DY^*_j\) to fit \(T_b\). During the estimation process, the hyper-parameters optimized are the number of variables randomly selected, \(N\), and the minimum number of observations to be kept in the terminal nodes of every fitted tree, \(Obs_{RF}\).\\

The second algorithm within the first level is Gradient Boosting (GB) with regression trees (\citeNP{Friedman_2000}). In this case, the gradient is minimized by sequentially fitting \(B\) regression trees. The subset of data to be used during the estimation process of every tree is also randomly selected. The expression to obtain the predicted scaled cumulative payments is
\begin{align}
\hat{D}^{*GB}_{ij}=\frac{\hat{D}^{GB}_{ij}}{P_{i}}=\hat{f}_{B-1}(X)+\delta_{GB} T_{B}(X)
\end{align}
\(\hat{f}_{B-1}(X)\) represents the function obtained after adding sequentially \(B-1\) regression tree models and, \(\delta_{GB}\) is the learning rate. The hyperparameter selected to be optimized during the training process is the minimum number of observations to be kept in the terminal nodes of every fitted tree, \(Obs_{GB}\). Regarding the hyperparameters, it is worth mentioning that the learning rate, \(\delta_{GB}\), is set to 0.01.\\

The last algorithm of the first layer is an Artificial Neural Network (ANN) (\citeNP{Mcculloch_1943}). Following the notation provided by \citeN{Bishop_2006} and taking into consideration that the feed-forward ANN used in this paper is composed of 2 hidden layers with 5 neurons each, the formal expression to obtain the predictions can be defined as follows:
\begin{align}
\begin{split}
&\hat{D}^{*ANN}_{ij}=\hat{D}^{ANN}_{ij}/P_{i}=\\
&=h^{(3)} \left( \sum_{k=1}^5 w_{1,k}^{(3)} h^{(2)} \left( \sum_{j=1}^5 w_{k,j}^{(2)} h^{(1)} 
 \left( \sum_{i=1}^2 w_{j,i}^{(1)} x_i + w_{j,0}^{(1)} \right) + w_{k,0}^{(2)} \right) + w_{1,0}^{(3)} \right)
\end{split}
\end{align}
where \(h^{(n)}\) is the activation function associated with layer \(n\), \(w_{z,v}^{(n)}\) is the \textit{v-th} weight associated with the neuron \(z\) inside layer $n$, and \(x_i\) refers to the \textit{i-th} input variable of the database composed of two explanatory variables, the scaled accident (\(AY^*_i\)) and development year (\(DY^*_j\)). The percentage of dropout regularization \(\theta\) is the hyperparameter to be optimized by applying a grid search and measuring the test error. As with the other algorithms, upper triangle predictions will be used as input within the second level of the architecture.\\

In addition to the three aforementioned algorithms, Chain Ladder assumptions are incorporated in the model architecture. To do so, the development factors of the Chain Ladder technique are used as an input in the second level of the Stacked-ANN model:
\begin{align}
\widehat{\lambda}^{*CL}_j=\frac{\sum^{n-j-1}_{i=1}{D^*_{ij}}}{\sum^{n-j-1}_{i=1}{D^*_{ij-1}}}
\end{align}
where \(\{\widehat{\lambda}^{*CL}_j: j=(2,3,\dots,J)\}\). Although the Chain Ladder methodology does not produce any parameters for \(j=1\), the second-level algorithm needs a value for \(j=1\). Thus, within the Stacked-ANN methodology, it is assumed that \(\widehat{\lambda}^{*CL}_1=1\).\\

Finally, CSR methodology (\citeNP{Meyers_2015}) is integrated. To achieve this, 10,000 MCMC simulations are produced within the first level of the Stacked-ANN model. Then, the expected scaled cumulative payments of the upper triangle arising from the aforementioned simulations are used as input in the algorithm within the second level of the Stacked-ANN model:
\begin{align}
\hat{D}^{*CSR}_{ij}=\frac{\sum^{10,000}_{k=1}{\hat{D}^{CSR}_{ijk}}/P_i}{10,000}
\end{align}

\subsection{Second level: Stacking algorithm}
\label{stack.second}
As previously stated, the inputs of this level are the scaled cumulative payments predicted by the algorithms named in Section \ref{stack.first} (RF, GB and ANN), the development factors based on the Chain Ladder technique and the expected scaled cumulative payments simulated by the CSR model. On the other hand, the output of the ANN within the second level and the Stacked-ANN are the cumulative payments \(\hat{D}^{*S-ANN}_{ij}\) by accident and development year.\\ 

\begin{figure}[!htb]
\begin{center}
\caption{Second-level structure}
\includegraphics[width=0.99\textwidth]{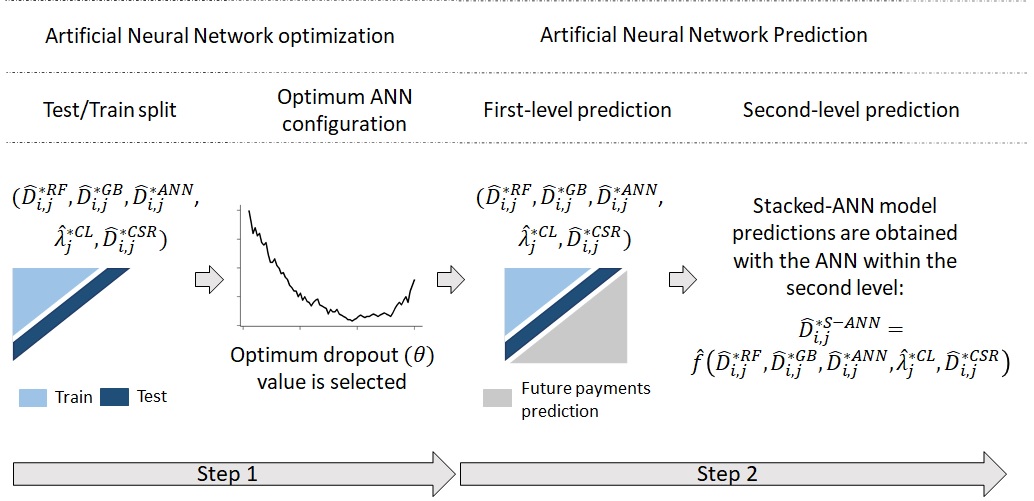}
\label{fig:Fig 4}
\end{center}
\end{figure}

Similar to the first-level algorithms, the training and optimization processes of the ANN within this level need to recognize that loss triangles are composed of a set of time series. The most recent information of the loss triangles is the last diagonal; thus, the explicative and response variables of this diagonal are selected as a test set, while the rest of the upper triangle data is used as a training set.\\

Once the test and training sets are defined, the optimum configuration of the ANN needs to be obtained. To do so, the training data are used to fit ANNs with different levels of dropout regularization \(\theta\). Then, the root mean squared error is computed by taking into consideration the predictions made by every ANN configuration. The \(\theta\) that minimizes the test error is selected.\\

Due to the Stacked-ANN architecture, two substeps need to be carried out in order to make the final predictions. First, the lower triangle of the first-level models need to be predicted. Second, the data predicted in the previous step are used as input of the ANN within the second layer to make the final predictions. Thus, the Stacked-ANN model tries to obtain more accurate predictions by combining different reserving models and algorithms.\\

Figure \ref{fig:Fig 1} shows the overall Stacked-ANN architecture, and Figure \ref{fig:Fig 4} provides a detailed summary of the process defined in the previous paragraphs. Technical details about the feedforward ANN fitted within this level of the Stacked-ANN model are presented below:
\begin{itemize}
\item It contains two hidden layers with 5 neurons each. The sigmoid activation function was selected for all neurons within the hidden layers while the linear activation function was used in the output layer, which is composed of one neuron.
\item The selected optimization algorithm is Adaptive Moment Estimation (ADAM), which was created by \shortciteN{dk_2014}. This method consists of a progressive adaptation of the initial learning rate, taking into consideration current and previous gradients. The default calibration proposed by the authors for the ADAM parameters is applied as \(\beta_1=0.9\) and \(\beta_2=0.999\). Thus, the ANN parameters are updated as follows:
\begin{gather}
\omega_t=\omega_{t-1}-\delta_{ANN}\frac{\hat{m}_t}{\sqrt{\hat{v}_t}+\epsilon}\\
\hat{m}_t=\frac{\beta_1 m_{t-1} + (1-\beta_1) g_t}{1-\beta_1^t}\\
\hat{v}_t=\frac{\beta_2 v_{t-1} + (1-\beta_2) g_t^2}{1-\beta_2^t}
\end{gather}
where \(\omega\) is the parameter to be updated and \(g_t\) the gradient in the epoch \(t\). The initial learning rate is set to \(\delta_{ANN}=0.01\).
\item The number of epochs is 10,000, and the batch size is equal to the length of the data used for training the ANN.
\item The backward pass calculations are done according to the selection of the root mean squared error as a loss function.
\item As previously stated, the percentage of dropout regularization \(\theta\) is the hyperparameter to be optimized by applying a grid search and measuring the test error.
\end{itemize}
Taking the abovementioned details into consideration, the scaled cumulative payments predicted by the Stacked-ANN model are obtained by means of the following expression:
\begin{align}
\begin{split}
&\hat{D}^{*S-ANN}_{ij}=\frac{\hat{D}^{*S-ANN}_{ij}}{P_{i}}=\widehat{f}(\hat{D}^{*RF}_{ij},\hat{D}^{*GB}_{ij},\hat{D}^{*ANN}_{ij},\widehat{\lambda}^{*CL}_j,\hat{D}^{*CSR}_{ij})=\\
&=h^{(3)} \left( \sum_{k=1}^{5} w_{1,k}^{(3)} h^{(2)} \left( \sum_{j=1}^{5} w_{k,j}^{(2)} h^{(1)} 
 \left( \sum_{i=1}^{5} w_{j,i}^{(1)} x_i + w_{j,0}^{(1)} \right) + w_{k,0}^{(2)} \right) + w_{1,0}^{(3)} \right)
\end{split}
\end{align}

\subsection{Log-normal simulation}
\label{stack.boot}
To compute the Kupiec test and the measures related to reserve variability (Section \ref{Benchmark}), the deviation around the central scenario predicted by the Stacked-ANN model needs to be obtained. Due to its right skewness and long tail, log-normal distribution is widely used within reserving models to derive the variability of the claims cost. Many papers used the lognormal distribution to compute this variability (see, among others, \citeN{Kremer_1982}, \shortciteN{abh_2006}, \citeN{rk_2009}, \citeN{wr_2013}, \citeN{Meyers_2015} or more recently, \shortciteN{onw_2018}).\\

In this study, a log-normal distribution is used to compute the reserve variability around the central scenario predicted by the Stacked-ANN. To do so, the parameters of this distribution are obtained using the aforementioned predictions and the moments method. Therefore, regardless of the distribution selected, the central scenario is that predicted by the Stacked-ANN, and thus, changing the distribution has no effect on the error measures described in Section \ref{Benchmark}. Nevertheless, changes to the log-normal hypothesis will modify the variability and, consequently, the risk measures (\(Ratio(RR^{t}_{1-\alpha})\) and \(Ratio(\sigma)\)) and the results of the Kupiec test. Below, the steps of the procedure are described:
\begin{enumerate}
\item Starting with the scaled cumulative payments predicted by the Stacked-ANN (\(\hat{D}^{*S-ANN}_{ij}\)), the variance by development year is computed as follows:
\begin{align}
Var[\hat{D}^{*S-ANN}_{j}]=\frac{\sum^{n}_{i=1}{\left(\hat{D}^{*S-ANN}_{ij}-E[\hat{D}^{*S-ANN}_{j}]\right)^2}}{n-1}
\end{align}
where \(n\) refers to the total number of accident years and \(E[\hat{D}^{*S-ANN}_{j}]\) is the mean of the scaled cumulative payments by development year.
\item By using the method of the moments and values calculated in the previous step, the parameters of the log-normal distribution are obtained:
\begin{align}
\widehat{\mu}_{ij}[\hat{D}^{*S-ANN}]&=\ln\left( \frac{E[\hat{D}^{*S-ANN}_{j}]^2}{\sqrt{Var[\hat{D}^{*S-ANN}_{j}]+E[\hat{D}^{*S-ANN}_{j}]^2}} \right)
\end{align}
\begin{align}
\widehat{\sigma}_j^2[\hat{D}^{*S-ANN}]&=\ln\left(1+\frac{Var[\hat{D}^{*S-ANN}_{j}]}{E[\hat{D}^{*S-ANN}_{j}]}\right)
\end{align}
\item For \(t=(1,2,...,T)\):
\begin{enumerate}
\item A triangle is generated by sampling random values from the following distribution function: \(\hat{C}^{*S-ANN,k}_{ij}\sim LN(\widehat{\mu}_{ij}[\hat{D}^{*S-ANN}],\widehat{\sigma}_j^2[\hat{D}^{*S-ANN}])\).
\item The final simulated values, \(\hat{C}^{S-ANN,k}_{ij}\), are obtained by removing the scaling. Hence, the scaled payments obtained in the previous step are multiplied by \(P_i\).
\end{enumerate}
\end{enumerate}

\section{Results}
\label{resul}
In this section, the data used, the fitting process and a final comparison between the Stacked-ANN and the benchmark models are shown.

\subsection{Data and fitting of the Stacked-ANN}
\label{data.fitting}
As stated in Section \ref{stack.data}, the upper and lower triangles required to fit and validate the models are obtained from Schedule P of the NAIC Annual Statement. This database contains the losses, reserves and premiums from 1988 until 1997 of different property and casualty insurers that underwrite business in the United States.\\

\citeN{Meyers_2015} indicated that one of the main mistakes with the NAIC Schedule P data is selecting triangles from insurers that made significant changes in their businesses. Meyers used the coefficient of variation of the net premiums and the net-on-gross ratio to select 50 triangles of each of the following lines of business: Commercial Auto (CA), Private Passenger Auto Liability (PA), Workers' Compensation (WC) and Other Liability (OL). This triangle selection was also used in this paper in order to ensure comparability with other studies that take as a reference the selection made by \citeN{Meyers_2015}. For further details about the data used to fit the Stacked-ANN, refer to Section \ref{stack.data}.\\

Once the data have been presented, the subsection focuses on the fitting of the Stacked-ANN. The first level of the proposed model is composed of three individual algorithms (RF, GB and ANN), the CSR reserving model and the development factors derived from the use of the Chain Ladder technique. The second level is composed of an ANN. As pointed out in Sections \ref{stack.first} and \ref{stack.second}, the optimum hyperparameters of the algorithms within the first and second levels are obtained for each triangle using a grid search. Table \ref{Hyper} lists the minor differences across the lines of business in the means of the 50 optimum hyperparameters obtained for each algorithm.\\

\begin{table}[H]
  \begin{center}
    \caption{Mean of the optimum hyperparameters by line of business}
    \label{Hyper}
    \begin{tabular}{l c c c c}
      \hline
      Line of     & RF first                  & GB first         & ANN first    &   ANN second        \\
      Business    & level                     & level            &  level       &   level             \\
      \hline
      CA          & $Obs_{RF}=2.04$; $N=1.94$ & $Obs_{GB}=4.38$  & $\theta=0.10$  & $\theta=0.09$     \\
      PA          & $Obs_{RF}=2.66$; $N=1.86$ & $Obs_{GB}=4.22$  & $\theta=0.07$  & $\theta=0.06$     \\
      WC          & $Obs_{RF}=1.78$; $N=1.68$ & $Obs_{GB}=3.66$  & $\theta=0.13$  & $\theta=0.12$     \\
      OL          & $Obs_{RF}=1.50$; $N=1.82$ & $Obs_{GB}=4.24$  & $\theta=0.12$  & $\theta=0.12$     \\
      \hline
     \multicolumn{2}{l}{\emph{Source}: own elaboration}    
    \end{tabular}
  \end{center}
\end{table}

As previously stated, the development factors (\(\widehat{\lambda}^{*CL}_j\)) obtained by applying the Chain Ladder technique to \(D^*_{ij}\) are used as input for the ANN included within the second level of the Stacked-ANN model. These values are calculated for each triangle. Table \ref{PaymentPattern} presents the means of the development factors by line of business.\\

With regard to the three algorithms of the first layer and the Chain Ladder technique, the CSR model is also incorporated in the Stacked-ANN architecture by means of inputting \(\hat{D}^{*CSR}_{ij}\) in the second-level algorithm. This Bayesian reserving model is fitted to every single triangle. Tables \ref{CSRParameter} and \ref{CSRParameterII} list the means of the CSR parameters by line of business.\\

\begin{table}[H]
  \begin{center}
    \caption{Mean of the development factors by line of business}
    \label{PaymentPattern}
    \begin{tabular}{c c c c c}
      \hline
      Development factors               & CA     & PA       & WC      & OL       \\
      \hline
      $\widehat{\lambda}^{*CL}_1$       & 1.89   &  1.77    &  2.21   &  6.66    \\ 
      $\widehat{\lambda}^{*CL}_2$       & 1.35   &  1.22    &  1.29   &  1.90    \\ 
      $\widehat{\lambda}^{*CL}_3$       & 1.16   &  1.10    &  1.13   &  1.33    \\ 
      $\widehat{\lambda}^{*CL}_4$       & 1.08   &  1.06    &  1.07   &  1.18    \\ 
      $\widehat{\lambda}^{*CL}_5$       & 1.04   &  1.03    &  1.04   &  1.10    \\ 
      $\widehat{\lambda}^{*CL}_6$       & 1.02   &  1.01    &  1.02   &  1.04    \\ 
      $\widehat{\lambda}^{*CL}_7$       & 1.00   &  1.01    &  1.02   &  1.02    \\ 
      $\widehat{\lambda}^{*CL}_8$       & 1.01   &  1.00    &  1.01   &  1.02    \\ 
      $\widehat{\lambda}^{*CL}_9$       & 1.00   &  1.00    &  1.01   &  1.01    \\ 
      $\widehat{\lambda}^{*CL}_{10}$    & 1.00   &  1.00    &  1.00   &  1.00    \\ 
      \hline
     \multicolumn{2}{l}{\emph{Source}: own elaboration}    
    \end{tabular}
  \end{center}
\end{table}

Table \ref{CSRParameter}, which is focused on the parameters needed to calculate the mean of the cumulative payments, presents positive \(\gamma\) and negative \(\beta_j\) for every line of business with the unique exception of CA, where \(\beta_6\), \(\beta_7\), \(\beta_8\) and \(\beta_9\) are positive. According to the model definition, the claims settlement speed increases when \(\beta_j<0\) and \(\gamma>0\). This common trend across the different lines of business about the claim settlement rate of the NAIC Schedule P data was already observed by \citeN{Meyers_2015}.\\

\begin{table}[H]
  \begin{center}
    \small
    \caption{CSR parameters by line of business: \(D_{i,j}\) mean}
    \label{CSRParameter}
    \begin{tabular}{c c c c c c c c c c c}
      \hline
      CSR             &          &          &          &           &  CSR             &          &          &          &          \\
      parameter       & CA       & PA       & WC       & OL        &  parameter       &  CA      &  PA      &  WC      &  OL      \\
      \hline
      $\alpha_1$      & 7.094    &  8.959   &  8.423   &  6.162    &  $\beta_1$       & -1.235   &  -0.987  &  -1.447  &  -2.446   \\ 
      $\alpha_2$      & 7.166    &  9.047   &  8.612   &  6.269    &  $\beta_2$       & -0.514   &  -0.400  &  -0.626  &  -1.332   \\ 
      $\alpha_3$      & 7.171    &  9.148   &  8.779   &  6.330    &  $\beta_3$       & -0.229   &  -0.198  &  -0.322  &  -0.709   \\ 
      $\alpha_4$      & 7.280    &  9.143   &  8.666   &  6.309    &  $\beta_4$       & -0.085   &  -0.097  &  -0.178  &  -0.363   \\ 
      $\alpha_5$      & 7.348    &  9.201   &  8.644   &  6.334    &  $\beta_5$       & -0.003   &  -0.042  &  -0.089  &  -0.173   \\ 
      $\alpha_6$      & 7.347    &  9.282   &  8.537   &  6.515    &  $\beta_6$       & 0.039    &  -0.017  &  -0.057  &  -0.079   \\ 
      $\alpha_7$      & 7.554    &  9.374   &  8.595   &  6.480    &  $\beta_7$       & 0.060    &  -0.008  &  -0.041  &  -0.045   \\ 
      $\alpha_8$      & 7.540    &  9.389   &  8.514   &  6.327    &  $\beta_8$       & 0.028    &  -0.001  &  -0.029  &  -0.030   \\ 
      $\alpha_9$      & 7.494    &  9.464   &  8.543   &  6.543    &  $\beta_9$       & 0.012    &  -0.001  &  -0.013  &  -0.014   \\ 
      $\alpha_{10}$   & 7.556    &  9.492   &  8.500   &  6.327    &  $\gamma$        & 0.021    &  0.008   &  0.016   &  0.028    \\  
      \hline
     \multicolumn{4}{l}{\emph{Source}: own elaboration}    
    \end{tabular}
  \end{center}
\end{table}

The comparison of the CSR deviation by development year presented in Table \ref{CSRParameterII} reveals that OL is the most volatile portfolio, while PA has the most stable reserves. For CA and WC, the reserve variability estimated by this Bayesian reserving model is quite similar, and it is located at an intermediate point between the OL and PA lines of business.\\

\begin{table}[H]
  \begin{center}
    \caption{CSR parameters by line of business: \(D_{i,j}\) STD}
    \label{CSRParameterII}
    \begin{tabular}{c c c c c}
      \hline
      CSR Parameter   & CA       & PA       & WC       & OL        \\
      \hline
      $\sigma_1$      & 0.303    &  0.028   &  0.236   &  0.771    \\ 
      $\sigma_2$      & 0.176    &  0.011   &  0.164   &  0.488    \\ 
      $\sigma_3$      & 0.109    &  0.007   &  0.117   &  0.327    \\ 
      $\sigma_4$      & 0.079    &  0.004   &  0.090   &  0.229    \\ 
      $\sigma_5$      & 0.063    &  0.003   &  0.069   &  0.164    \\ 
      $\sigma_6$      & 0.052    &  0.002   &  0.052   &  0.120    \\ 
      $\sigma_7$      & 0.043    &  0.002   &  0.037   &  0.087    \\ 
      $\sigma_8$      & 0.035    &  0.001   &  0.025   &  0.061    \\ 
      $\sigma_9$      & 0.026    &  0.001   &  0.016   &  0.038    \\ 
      $\sigma_{10}$   & 0.014    &  0.001   &  0.008   &  0.019    \\  
      \hline
     \multicolumn{2}{l}{\emph{Source}: own elaboration}    
    \end{tabular}
  \end{center}
\end{table}

\subsection{Comparison against benchmark models}
\label{Comparison}
Once the Stacked-ANN reserving model is fitted, its performance is compared with the benchmark models explained in Section \ref{Benchmark} (ODP, Mack, CSR and an individual ANN).\\

Table \ref{RMSE} lists the \%RMSEs associated with reserves \(R^{t}\), next year payments \(P^{t+1}\) and ultimate losses \(U^{t}\) by line of business and reserving model. For further details about the measures presented in the table, refer to Section \ref{Benchmark}. \\

\begin{table}[H]
  \begin{center}
    \caption{\%RMSE by line of business and reserving model}
    \label{RMSE}
    \begin{tabular}{l c c c c c c}
      \hline
      Error              & Line of  & ODP       & Mack's    & CSR       & ANN       & Stacked  \\
      Measure            & Business &           & Model     &           &           & ANN      \\
      \hline
      $\%RMSE(R^{t})$    & CA       & 0.896\%   & 0.896\%   & 0.534\%   & 1.768\%   & 0.739\%  \\ 
      $\%RMSE(P^{t+1})$  & CA       & 0.668\%   & 0.669\%   & 0.573\%   & 1.775\%   & 0.876\%  \\ 
      $\%RMSE(U^{t})$    & CA       & 0.170\%   & 0.171\%   & 0.102\%   & 0.337\%   & 0.141\%  \\
      \hline
      $\%RMSE(R^{t})$    & PA       & 1.012\%   & 1.004\%   & 0.823\%   & 5.006\%   & 0.254\%  \\ 
      $\%RMSE(P^{t+1})$  & PA       & 1.290\%   & 1.286\%   & 0.258\%   & 1.900\%   & 0.320\%  \\ 
      $\%RMSE(U^{t})$    & PA       & 0.131\%   & 0.131\%   & 0.107\%   & 0.651\%   & 0.033\%  \\
      \hline 
      $\%RMSE(R^{t})$    & WC       & 1.295\%   & 1.286\%   & 1.751\%   & 1.943\%   & 1.058\%  \\ 
      $\%RMSE(P^{t+1})$  & WC       & 0.887\%   & 0.880\%   & 1.531\%   & 1.525\%   & 0.676\%  \\ 
      $\%RMSE(U^{t})$    & WC       & 0.222\%   & 0.221\%   & 0.301\%   & 0.333\%   & 0.182\%  \\
      \hline 
      $\%RMSE(R^{t})$    & OL       & 5.274\%   & 5.086\%   & 3.153\%   & 5.725\%   & 0.722\%  \\ 
      $\%RMSE(P^{t+1})$  & OL       & 2.216\%   & 2.102\%   & 5.528\%   & 0.268\%   & 1.095\%  \\ 
      $\%RMSE(U^{t})$    & OL       & 1.760\%   & 1.709\%   & 1.056\%   & 1.918\%   & 0.242\%  \\
      \hline 
     \multicolumn{2}{l}{\emph{Source}: own elaboration}    
    \end{tabular}
  \end{center}
\end{table}

The results obtained by using the different reserving models are summarized as follows:
\begin{itemize}
\item The Stacked-ANN model outperforms the individual ANN. The proposed architecture is empirically more accurate because it can learn from the reserving models (Chain Ladder and CSR) and machine learning algorithms (RF, GB and ANN) included within the first level of the Stacked-ANN, while the individual ANN must base its training only on the origin data (\(AY^*_i\) and \(DY^*_j\)) without taking advantage of other models that are able to capture different patterns and characteristics.
\item As they are based on Chain Ladder assumptions, the mean of the distributions generated by ODP and Mack's model should converge to the values obtained by applying the deterministic approach of the Chain Ladder technique. Consequently, the error measures observed in Table \ref{RMSE} for these two stochastic reserving models are almost the same. The table also reveals that ODP and Mack are less accurate than the Stacked-ANN model in most cases. \(\%RMSE(P^{t+1})\) of CA is a unique category in which the benchmark models based on the Chain Ladder technique are more accurate than the proposed methodology.
\item Regarding the comparison between Stacked-ANN and CSR, \(R^{t}\) and \(U^{t}\) of PA, WC and OL estimated by the proposed model are significantly more accurate than those obtained when using the Bayesian model. Additionally, \(\%RMSE(P^{t+1})\) of the Stacked-ANN model is lower in WC and OL. Thus, in the majority of cases, the CSR model is outperformed by the proposed methodology.
\end{itemize}

\begin{figure}[!htb]
\begin{center}
\caption{\(\%RMSE(R^{t})\) by line of business and volume of reserves.}
\includegraphics[width=0.99\textwidth]{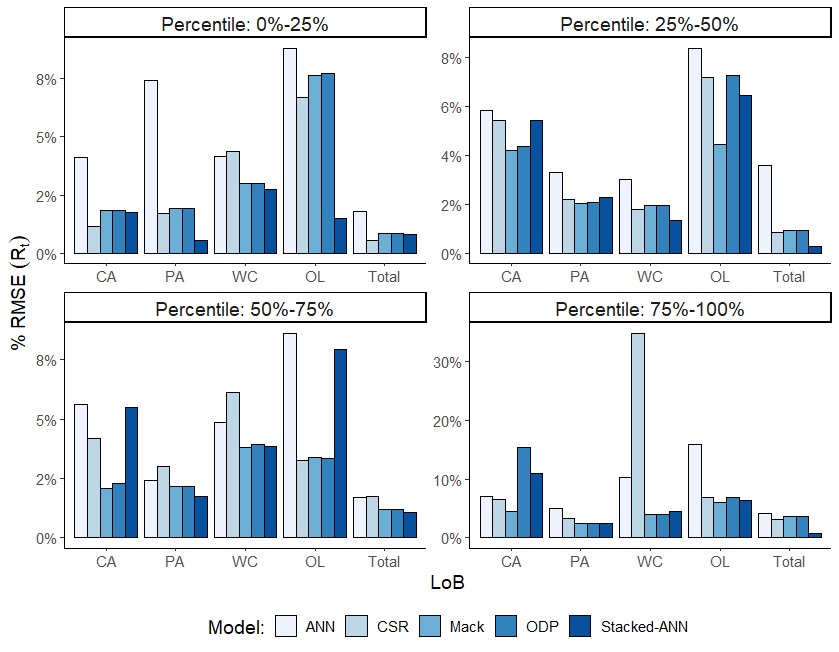}
\label{fig:Fig 2}
\end{center}
\end{figure}

To enhance the analysis of the results presented in Table \ref{RMSE}, Figure \ref{fig:Fig 2} shows the \(\%RMSE(R^{t})\) by line of business and volume of reserves. First, the companies were classified in four different groups taking into consideration the volume of reserves and the quartiles associated with the distribution. Then, the error rate of each reserving model was computed by line of business. The former calculation was carried out without making any distinction between lines of business.\\

The results of the aforementioned figure reveal that when no distinction between lines of business is made, the Stacked-ANN architecture outperforms the rest of the reserving models regardless of the company size. This difference is especially relevant for those companies with a higher level of reserves, whose results are collected in the graph labelled `Percentile: 75\%-100\%'. As expected, some fluctuations in the performance of the models are observed when the results are analysed by line of business and volume of reserves. Nonetheless, the error rate of the Stacked-ANN tends to be lower than the rest of the benchmark models.\\

In accordance with the reasons explained within the former paragraphs, it can be concluded that the Stacked-ANN model takes advantage of the different characteristics of several reserving models and machine learning algorithms, leading to a more flexible and precise architecture in most of the cases.\\

In addition to the error analysis, the risk measures (\(Ratio(RR^{t}_{1-\alpha})\) and \(Ratio(\sigma)\)) and the p-values of the Kupiec test obtained by using each reserving model are compared. Before examining the results, it is important to point out that Mack's model does not make any assumptions about the payment distribution, ODP assumes that incremental payments follow an overdispersed Poisson distribution, and CSR, ANN and Stacked-ANN presume that cumulative payments are log-normally distributed. The hypothesis taken regarding the payments impact the distribution shape and consequently the risk measures. Therefore, in this case, ODP and Mack's model are not going to converge like they did in the central scenario.\\ 

\begin{table}[!htbp]
  \begin{center}
    \caption{Risk measures by line of business and reserving model}
    \label{Risk}
    \begin{tabular}{l c c c c c c}
      \hline
      Risk                    & Line of  & ODP         & Mack's      & CSR         & ANN         & Stacked    \\
      measures                & Business &             & Model       &             &             & ANN        \\
      \hline
      $Ratio(RR^{t}_{0.995})$ & CA       & 1.936       & 1.460       & 2.776       & 1.387       & 1.884       \\ 
      $Ratio(\sigma)$         & CA       & 2.561       & 0.461       & 0.681       & 0.456       & 0.642       \\ 
      Kupiec p-value          & CA       & $\geq 0.05$ & $\geq 0.05$ & $\geq 0.05$ & $<0.05$     & $\geq 0.05$ \\
      \hline
      $Ratio(RR^{t}_{0.995})$ & PA       & 0.544       & 0.373       & 0.918       & 0.783       & 0.888       \\
      $Ratio(\sigma)$         & PA       & 0.277       & 0.135       & 0.270       & 0.279       & 0.332       \\
      Kupiec p-value          & PA       & $\geq 0.05$ & $\geq 0.05$ & $\geq 0.05$ & $\geq 0.05$ & $\geq 0.05$ \\
      \hline
      $Ratio(RR^{t}_{0.995})$ & WC       & 2.525       & 0.691       & 1.797       & 2.194       & 2.149       \\ 
      $Ratio(\sigma)$         & WC       & 1.273       & 0.245       & 0.474       & 0.682       & 0.717       \\ 
      Kupiec p-value          & WC       & $<0.05$     & $<0.05$     & $<0.05$     & $<0.05$     & $\geq0.05$  \\
      \hline
      $Ratio(RR^{t}_{0.995})$ & OL       & 7.506       & 3.287       & 4.843       & 2.315       & 3.522       \\ 
      $Ratio(\sigma)$         & OL       & 6.275       & 1.217       & 1.119       & 0.690       & 1.099       \\ 
      Kupiec p-value          & OL       & $\geq 0.05$ & $\geq 0.05$ & $\geq 0.05$ & $\geq 0.05$ & $\geq 0.05$ \\
      \hline
     \multicolumn{2}{l}{\emph{Source}: own elaboration}    
    \end{tabular}
  \end{center}
\end{table}

The \(Ratio(RR^{t}_{1-\alpha})\) and the p-values collected in Table \ref{Risk} evaluate the \(99.5\) percentile (\(\alpha=0.005\)) of the reserve distribution function, which is the confidence level set up by Solvency II to calculate the risk of the insurance companies. The results of this table are summarized below:
\begin{itemize}
\item According to the results of the Kupiec test, the Stacked-ANN generates an adequate risk assessment for every line of business. It is worth mentioning that when compared with the individual ANN, the empirical results show that the stacking process not only improves the error rate but also allows for the generation of more appropriate distribution functions using the same simulation approach (presented in Section \ref{stack.boot}). With regard to the comparison between the Stacked-ANN and the rest of benchmark models, the Kupiec test reveals that CSR, ODP and Mack's model do not produce appropriate risk measures for WC, while the proposed methodology passes the test.
\item Intuitively, the duration of the liabilities should have a close relation with \(Ratio(RR^t_{0.995})\) and \(Ratio(\sigma)\): the longer the duration, the higher is the uncertainty around each economic unit of reserve. The development factors based on the Chain Ladder technique measure the claim settlement speed. Therefore, they can be considered a good indicator of the duration of liabilities. A development factor at year \(t\), \(\lambda_t\), means that the \(t+1\) cumulative payment is \(\lambda_t\) times the cumulative claims settled at \(t\). Consequently, high development factors indicate a long duration, while low values reflect a high settlement speed. According to Table \ref{PaymentPattern}, OL is the line of business with the highest duration, while PA has the lowest. CA and WC, whose durations are in a similar range, are located at an intermediate point between PA and OL. As can be observed in Table \ref{Risk}, this intuition about the relation between the duration and reserve uncertainty is followed by the Stacked-ANN and benchmark models.
\item In general, the \(Ratio(RR^t_{0.995})\) and \(Ratio(\sigma)\) by line of business are similar across the different reserving models. The two main exceptions are the risk measures of ODP for OL and Mack for WC. The high values observed in the ODP estimations for OL are due to two companies whose \(RR^{t}_{1-\alpha}/\hat{R}^{t}_{k,\mu}\) ratios are higher than \(60\), while in the second case, Mack's model systematically underestimates the variability of the payments, leading to lower values compared with the rest of the models and an inadequate risk assessment according to the results of the Kupiec test. The \(Ratio(RR^t_{0.995})\) and \(Ratio(\sigma)\) of the Stacked-ANN are in line with the majority of the benchmark models, and no extremely high/low risk measures are observed in Table \ref{Risk}.
\end{itemize}

\subsection{Sensitivity analysis of the number of hidden layers}
\label{Sensitivity}
As explained in Section \ref{stack}, the ANNs included within the proposed Stacked-ANN architecture are composed of two hidden layers, each with five neurons. To analyse the impact of the ANN complexity (\citeNP{Cyb_1989}, \shortciteNP{HSW_1989}, \citeNP{Hor_1991} and \shortciteNP{LLP_1993}, among others, introduced the theoretical framework to analyse the approximation capabilities of neural networks) on the predictive power of the Stacked-ANN model, a sensitivity analysis of the number of hidden layers was carried out. Thus, Table \ref{Sensitivity} compares the configuration selected for the Stacked-ANN model in this paper with two alternative configurations: ANNs composed of one and three hidden layers with five neurons each.\\

\begin{table}[!htbp]
  \begin{center}
    \caption{Sensitivity analysis of the number of hidden layers}
    \label{Sensitivity}
    \begin{tabular}{c c c c c c c}
      \hline
      Hidden  & Line of  &                 &                   &                 \\
      layers  & Business & $\%RMSE(R^{t})$ & $\%RMSE(P^{t+1})$ & $\%RMSE(U^{t})$ \\
      \hline
      1       & CA       & 0.840\%         & 0.766\%           & 0.160\%         \\ 
      2       & CA       & 0.739\%         & 0.876\%           & 0.141\%         \\ 
      3       & CA       & 0.780\%         & 0.643\%           & 0.149\%         \\
      \hline
      1       & PA       & 2.512\%         & 3.507\%           & 0.326\%         \\
      2       & PA       & 0.254\%         & 0.320\%           & 0.033\%         \\
      3       & PA       & 0.231\%         & 0.115\%           & 0.030\%         \\
      \hline
      1       & WC       & 1.398\%         & 1.296\%           & 0.240\%         \\ 
      2       & WC       & 1.058\%         & 0.676\%           & 0.182\%         \\ 
      3       & WC       & 1.140\%         & 1.030\%           & 0.196\%         \\
      \hline
      1       & OL       & 1.419\%         & 1.145\%           & 0.475\%         \\ 
      2       & OL       & 0.722\%         & 1.095\%           & 0.242\%         \\ 
      3       & OL       & 0.613\%         & 0.794\%           & 0.205\%         \\
      \hline
     \multicolumn{2}{l}{\emph{Source}: own elaboration}    
    \end{tabular}
  \end{center}
\end{table}

Two main conclusions can be drawn from the results obtained. First, the high level of error of the one hidden layer model demonstrates that more complexity is needed in order to properly predict general insurance reserves. The structure proposed during this study for the Stacked-ANN model (two hidden layers) performs significantly better than this first alternative in every single line of business.\\

Second, the performance of the three hidden layers alternative is similar to that of the suggested architecture. As no significant differences are observed, the two hidden layer structure is considered more appropriate because the three hidden layer structure adds complexity to the model without a significant improvement in the error rate.\\

\section{Conclusions}
\label{conc}
This paper introduced a stochastic reserving model based on stacking different machine learning algorithms (RF, GB and ANN) and reserving models (Chain Ladder and CSR). The predictive power and reserve volatility of the proposed approach, named Stacked-ANN, were compared with stochastic reserving models based on the Chain Ladder technique (ODP and Mack's model), an individual ANN and CSR, which is a Bayesian loss reserving model.\\

Three main conclusions were drawn. First, a comparison of the Stacked-ANN with the individual ANN revealed that the predictions of the reserves \(R^{t}\), next year payments \(P^{t+1}\) and ultimate losses \(U^{t}\) made by machine learning algorithms were improved by applying the proposed stacking procedure. The hybrid architecture learns patterns and characteristics from several algorithms and reserving models, resulting in a more flexible and accurate model than an individual ANN, whose inputs for training are limited to the original data.\\

Second, the empirical results indicated that the Stacked-ANN model is more precise than CSR and the most widely used stochastic reserving models based on the Chain Ladder technique (ODP and Mack's model). In particular, the \(R^{t}\) and \(U^{t}\) predictions made by the Stacked-ANN were more precise than those of ODP and Mack's model in all the lines of business analysed, while the Bayesian model (CSR) was outperformed by the proposed architecture in three out of four lines of business. It is important to remark that in Other Liability (OL), which is a line of business with a longer duration and therefore a portfolio where the importance of an accurate reserves estimation is especially relevant, the error of the models based on Chain Ladder or Bayesian statistics was more than four times the error of the Stacked-ANN. Therefore, it can be concluded that machine or deep learning techniques can be used to improve the performance of the traditional reserving techniques based on Bayesian statistics or the Chain Ladder.\\

With regard to accuracy, it is worth mentioning that the proposed structure of the ANNs (two hidden layers) within the Stacked-ANN model seems to be the optimal configuration according to the empirical results. On the one hand, the error increased significantly when the number of hidden layers is reduced to one. On the other hand, the results demonstrated that increasing the number of hidden layers does not have an impact on the accuracy. Thus, increasing the complexity of the ANNs by up to three hidden layers will extend the training phase without making any significant improvement in the error.\\

Third, the results of a Kupiec test revealed that the risk estimation made by the Stacked-ANN can be considered as appropriate in all lines of business analysed, while the rest of the benchmark models failed the test at least once. In particular, CSR, ODP and Mack's model were unable to produce an appropriate p-value for the Kupiec test in the Workers' Compensation (WC) business, while the individual ANN failed the test in Commercial Auto (CA) and, as with the previous models, in Workers' Compensation. Taking into consideration that the same log-normal approach was used to obtain the reserves variability of the individual ANN and the Stacked-ANN, it must be mentioned that the stacking procedure not only increases the accuracy but also allows for the simulation of more adequate distribution functions.\\

The aforementioned robustness and predictive power of the Stacked-ANN compared with other reserving models suggest that further investigation should be conducted about the possible application of this model within the \textit{actuary in the box} approach. The generation of outliers is one of the main problems when using the former methodology with Chain Ladder models. Therefore, the robustness of the Stacked-ANN can be exploited in order to improve the \textit{actuary in the box} methodology, which is widely used to assess the fact that reserves can be insufficient to cover their runoff over a 12-month time horizon.\\

\bibliography{Final1}
\bibliographystyle{chicago}
\end{document}